\documentclass[pra,twocolumn,showpacs,superscriptaddress]{revtex4-1}
\def\iota{\imath}
\newcommand{\bfsfG}{\mbox{\sffamily\bfseries{G}}}
\usepackage[psamsfonts]{amssymb}
\usepackage{amsmath}
\usepackage{epsf,epsfig}
\usepackage{graphicx}
\usepackage{lipsum}
\begin{document}
\title{Entanglement dynamics of two two-level atoms in the vicinity of an invisibility cloak\\}
\author{Ehsan Amooghorban}
\email{Ehsan.amooghorban@sci.sku.ac.ir} \affiliation{Department of
Physics, Faculty of Basic Sciences, Shahrekord University, P.O. Box
115, Shahrekord 88186-34141, Iran.} \affiliation{Photonics Research
Group, Shahrekord University, Shahrekord 88186-34141,
Iran.}
%
%
\author{Elnaz Aleebrahim}
\affiliation{Department of Physics, University of Isfahan, Hezar Jarib Avenue, Isfahan, Iran.}

\begin{abstract}
We study the entanglement between two identical two-level atoms located near an ideal model of invisibility cloaks, by monitoring
the the time evolution of the concurrence measure.
We obtain the reduced density operator of the atomic subsystem based on a canonical quantization scheme presented for the electromagnetic field interacting with atomic systems in the presence of an anisotropic, inhomogeneous and absorbing magnetodielectric medium. It is shown that two atoms, which are prepared initially in an unentangled state, are correlated in weak coupling regime via the spontaneous emission and the dipole-dipole interaction of two atoms mediated by the invisibility cloak. We therefore find that the invisibility cloak, independent of the hidden object, works fairly well at frequencies far from the resonance frequency of the object and the cloak, whereas near the resonance frequency, the hidden object becoming detectable due to a sharp reduction of the concurrence.
\end{abstract}

\maketitle

\section{Introduction}
Recent progress in the development of nanofabrication techniques has opened the possibility to create artificial media with
subwavelength building blocks have electromagnetic (EM) responses that give rise to fascinating effects such as negative refraction~\cite{Pendry 2000,Luo 2002,Pendry 2004,Smith 2006}, artificial magnetism~\cite{Pendry 1999,Brien 2002,Yen 2004}, perfect absorbers~\cite{Landy 2008} and hyperbolic dispersion~\cite{Jacob 2006,Liu 2007}.
Such media, known as metamaterials, allow unprecedented
control over the propagation of EM fields, which are not available to natural materials. Controlling the EM waves at will has enabled designing a novel device that deflect light around a cloak and render objects inside invisible. This idea, which originally proposed by Pendry {\rm et al.}~\cite{Pendry 2006,Leonhardt 2006} based on transformation optics, became more evident with the successful experimental realizations of the cloaks at the microwave and optical frequencies~\cite{Schurig 2006,Cummer 2006,Cai 2007} and the carpet cloaking~\cite{Li 2008}. Such spatial cloaks have also offered new attractive opportunities in the contexts of acoustic cloaking~\cite{Urzhumov 2010,Zhang 2011,Sanchis 2013}, and heat cloaking~\cite{Muhammad 2016}.

Recently, a lot of attempts have been made to study the invisibility cloaks in the framework of quantum mechanic,
and becomes an important research topic. While efforts in this field are now mainly directed at the quantum cloaking of matter waves~\cite{Zhang 2008,Lin 2009,Lin 2011,Chen 2012,Chang 2014,Fleury 2014}, which requires a proper design of the effective mass and potential of the cloaking shell, there is other interesting possibility in the context of the EM cloak by treating light as a stream of photons rather than as electromagnetic waves. Prime motivations for this come from studying the propagation of quantum states in metamaterials. For active and some passive metamaterials, it is shown that the classical effective parameters are insufficient and an additional
effective-medium parameter is needed to correctly predict the properties of the quantum states of light
emerging from the metamaterials~\cite{Amooghorban 2013,Amooghorban arXiv}.

Inspired by this recent work, the spontaneous decay and its spatial distribution of the spontaneously emitted light, which is a fundamental quantum property of emitters, has been studied for an excited atom close to a spherical invisibility cloak~\cite{Morshed 2016}.
Since the radiative properties of an excited atom depends on both the atom and the material environment, the relaxation to the ground state via spontaneous emission is surely modified by the cloaked object. On other hand, it is expected that the radiation dynamics of a source in the presence of the cloaking device differ drastically from what experiences in the presence of natural materials, because of unusual properties of such devices. Hence, the
spontaneous decay was used as a probe to study the effect of the EM cloak from a different perspective, which is absent in a classical treatment.
It was found that the spontaneous emission rate can be either enhanced or suppressed sharply by the invisibility cloak. In this sense, this
was also an appropriate tools to determine the presence of the cloak and any objects it contained even in the weak-coupling regime~\cite{Morshed 2016}, but the material composition inside the cloak were not detectable. It means that the clock behaved more like a super-visibility device.

This was also the subject of study by a number of other groups over the past several years. The spontaneous emission rate of an excited two level atom and its radiative dynamics is studied inside a plasmonic cloak~\cite{Rahmani 2013,Kamp 2013}. The possibility to control and manipulate the spontaneous emission rate of an atom with different devices designed with transformation optics is investigated in~\cite{Zhang 2016}.

The quantum entanglement is an another striking feature of quantum mechanics, which has no counterpart in classical mechanics. It demonstrates the non-local character of quantum mechanics and is the key physical resource in most quantum information processes such as quantum cryptography, quantum teleportation, and quantum computation~\cite{Bennett 1993,Grover 1999,Nielsen 2000}. Systems of two-level atoms interacting with the fields have been one of the important prototypes in the investigations concerning entanglement.
It is also known that an environment usually leads to decoherence and noise, which may
cause entanglement disappear and even in certain circumstances enhancement~\cite{Braun 2002,Kim 2002,Jak´obczyk 2002,Schneider 2002,Benatti 2003}.
It is useful to look at the entanglement dynamics of a quantum system in vicinity of a spherical invisibility cloak as a probe to gain new understanding of the EM cloak effect from a quantum perspective.
We assume that the quantum system composed by two two-level atoms initially prepared in a separable state and symmetrically placed at fixed radial distances outside the cloaked object and coupled to the EM field.

In the absence of invisibility cloak and the hidden object, two atoms are entangled via the spontaneous emission and dipole-dipole interaction~\cite{Tanas 2003,Tanas 2004}.
Based on a classical treatment, the combination of the cloak and the object have the properties of free space when viewed externally.
It is of interest to see whether the dynamical behavior of entanglement for aforementioned atomic system differs drastically from what would experience in free space, of course, if the cloak can hide the object successfully.
This is what we are going to pursue in the present paper.
In the following, we consider a canonical quantization approach~\cite{Huttner 1992,Jeffers1996,Suttorp2004,Amooshahi 2009,Kheirandish 2010,Philbin 2010,Kheirandish 2011,Amooghorban 2011,Amooghorban 2015} to investigate the entanglement dynamics of the atomic subsystem in the presence of the cloaking devices.

This paper is organized as follows. In Sec.~\ref{Sec:The basic relations}, we present a brief overview on the canonical quantization of EM field interacting with charged particles in the presence of an isotropic, homogeneous, and absorbing magnetodielectric (MD) medium.
We apply this approach to the case of two atoms are at an equal distance from a cloaked object.
The Hamiltonian governing the interaction between the atomic system, the cloaked object and the EM field is obtained in the electric-dipole and the rotating-wave approximation.
Then, the evolution of the system is determined by the Schrodinger equation.
In Sec.~\ref{Sec:The basic relations}, with the suppose of weakly interacting with the EM field, the
reduced operator of the two two-level atoms is obtained from the complete time evolution describing the total system by tracing over
the field degrees of freedom.
Analytical results are presented for concurrence, which is well known and calculable measure of entanglement.
We employ the Lorentz model to describe the material absorption and dispersion of the cloaked object.
Then, by means of the concurrence the amount of entanglement created in the atomic system is numerically calculated during the spontaneous emission.
Our conclusions are summarized in Sec.~\ref{Sec:conclusion}. In Appendix~\ref{App:Details of canonical quantization}, we present briefly some details of the canonical quantization of EM field in the presence of media. The details of the Green's tensor evaluation can be found in Appendix~\ref{App:GREEN TENSOR OF THE SYSTEM}.
%
%
\section{The basic relations}\label{Sec:The basic relations}
\begin{figure}[t]
\includegraphics[width=0.7\columnwidth]{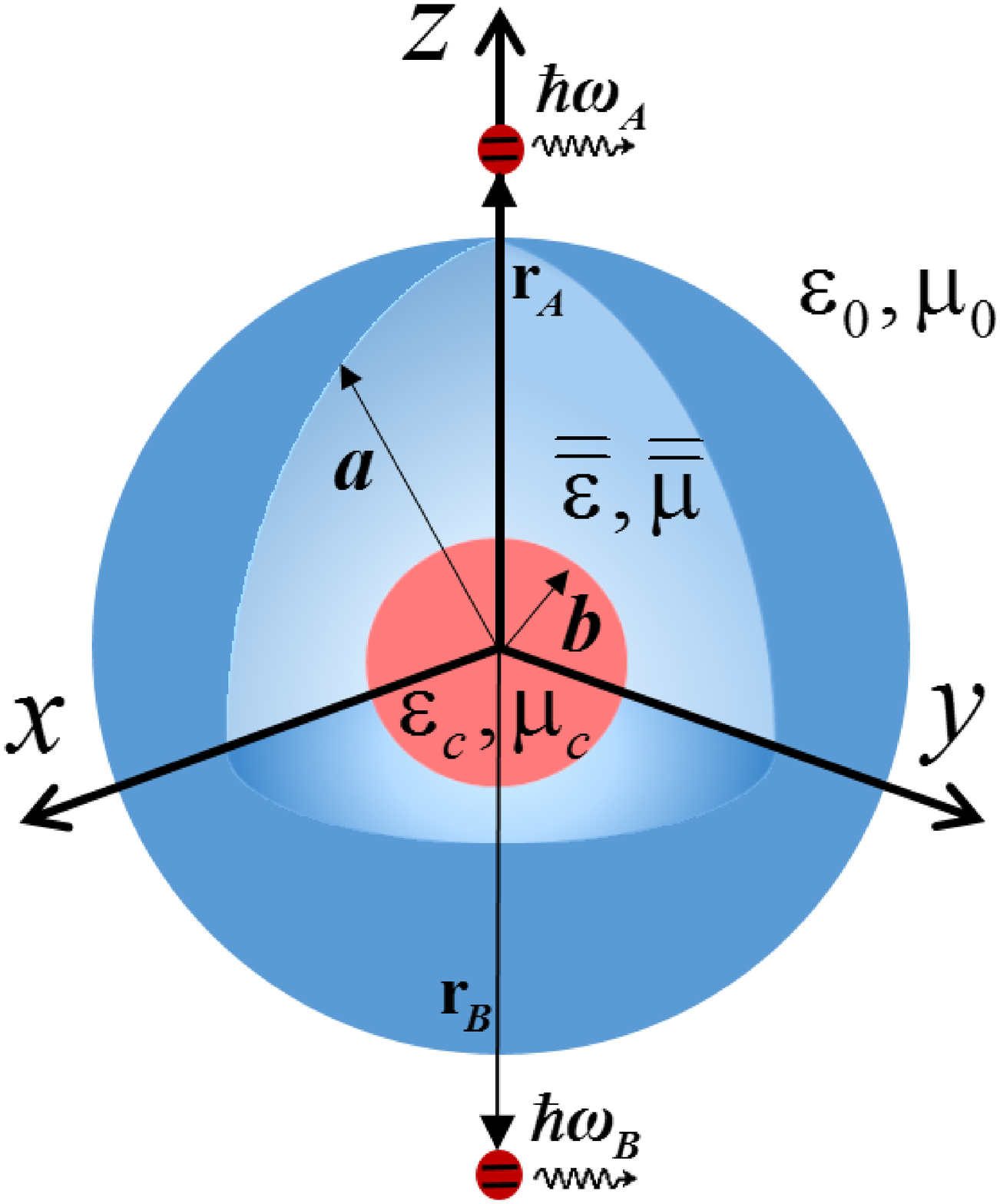}
\caption{The geometry of the system. Two two-level atoms are placed at fixed points ${\bf r}_a$ and ${\bf r}_b$ near a spherical invisibility cloak with the material parameters $\overline{\overline{\boldsymbol \varepsilon }}$ and $\overline{\overline{\boldsymbol \mu }}$ are given by Eq.~(\ref{material parameters of the cloak}). The clock has inner and outer radius of $b$ and $a$, respectively, and
a homogeneous and isotropic material is inserted to the central hollow region of the cloak, $r < b$, with the electric primitivity and the magnetic permeability function
$\varepsilon_c$ and $\mu_c$.}\label{Fig:1}
\end{figure}

We consider a pair of atoms, modeled by two two-level atoms with the dipole moments ${\bf d}_{A}$ and ${\bf d}_{B}$ and the transition frequencies ${{\omega }_{A}}$ and ${{\omega }_{B}}$. These are located symmetrically near an ideal invisibility cloak and coupled to the vacuum field. A sketch of the system under study can be seen in Fig.~\ref{Fig:1}. Due to the spherical symmetry of the cloaking device, without loss the generality,
we specify the atomic system such that two atoms are placed on the $z$-axis outside the cloak at the positions ${{\mathbf{r}}_{A}}$ and ${{\mathbf{r}}_{B}}$.
The ideal spherical invisibility cloak can be designed by considering a spherically symmetric transformation optics, that maps a spherical region $0<r<a$
into an annular region $b<r<a$. Based on this transformation, the material parameters of the cloak are given by~\cite{Pendry 2006}
\begin{subequations}\label{material parameters of the cloak}
\begin{eqnarray}
\overline{\overline{\boldsymbol \varepsilon }}\left( \mathbf{r},\omega  \right)={{\varepsilon }_{0}}\left[ \left( {{\varepsilon }_{r}}-{{\varepsilon }_{t}} \right) \right]\hat{\mathbf{r}}\hat{\mathbf{r}}+{{\varepsilon }_{t}}\bar{\bar{\bf I}}, \\
\overline{\overline{\boldsymbol \mu }}\left( \mathbf{r},\omega  \right)={{\mu }_{0}}\left[ \left( {{\mu }_{r}}-{{\mu }_{t}} \right) \right]\hat{\mathbf{r}}\hat{\mathbf{r}}+{{\mu }_{t}}\bar{\bar{\bf I}},
\end{eqnarray}
\end{subequations}
where $\overline{\overline{\bf I}}=\hat{\bf r}\hat{\bf r}+\hat{\boldsymbol \theta}\hat{\boldsymbol \theta}+\hat{\boldsymbol \phi}\hat{ \boldsymbol \phi}$ is the unit dyad. Here, the subscripts r and t refer to the parameters along radial $\hat{{\bf r}}$ and tangential directions $\hat{\boldsymbol \theta }$ and $\hat{\boldsymbol \phi }$, respectively, and the permittivity and permeability tensor components for the cloaking shell are given by
\begin{subequations}
\begin{eqnarray}
&& {{\varepsilon }_{r}\left( \mathbf{r},\omega  \right)}={{\mu }_{r}\left( \mathbf{r},\omega  \right)}= \frac{{b}{{\left( r-{a} \right)}^{2}}}{({b}-{a}){r}^{2}}{{\kappa }_{L}\left( \omega  \right)}, \\
&& {{\varepsilon }_{t}\left( \omega  \right)}={{\mu }_{t}\left( \omega  \right)}=\frac{{b}}{({b}-{a})}{{\kappa }_{L}\left( \omega  \right)}.
\end{eqnarray}
\end{subequations}
It is well known that the cloaking devices are unavoidably frequency dependent (i.e., dispersive)~\cite{Pendry 2006}.
In reality, finite dissipation accompanies by absorption to fulfill
causality.
Based on this dissipative property, the permittivity and permeability tensor components of the invisibility cloak are multiplied by a Lorentzian factor, ${{\kappa}_{L}}(\omega )= 1+\omega _{p}^{2}/(\omega _{0}^{2}-{{\omega }^{2}}-i\gamma \omega ) $~\cite{Jackson 1999}, where in which ${{\omega }_{p}}$ and ${{\omega }_{0}}$ are respectively plasma frequency and resonance frequency and $\gamma $ is the absorbtion coefficient of the invisibility cloaking. For simplicity, an isotropic and homogeneous MD medium is considered to specify the inserted object in the central hole of the cloak. The material parameters of the hidden object are $\varepsilon \left( \omega  \right)=\mu \left( \omega  \right)=\alpha {{\kappa}_{L}}(\omega )$ where $\alpha $ is a constant.

The quantized EM field interacting with the atomic system in the presence of lossy and dispersive cloaked object should be described in the framework of the second quantization.
Generally, the quantization scheme of the EM field in the presence of absorbing and dispersive media has been carried out in two approaches: Canonical~\cite{Huttner 1992,Jeffers1996,Suttorp2004,Amooshahi 2009,Kheirandish 2010,Philbin 2010,Kheirandish 2011,Amooghorban 2011,Amooghorban 2015} and phenomenological methods~\cite{Gruner 1996,Dung 1998,Scheel 1998,Matloob 1995,Knoll 2001,Dung 2000,Dung 2003,Matloob 2004}. In the present paper, we use the rigorous canonical formalism to gain a better understanding of the mechanism of interaction in a microscopic level.
Based on this microscopic approach, the quantization of the EM field in an inhomogeneous,
anisotropic, dissipative, and dispersive MD
medium can be accomplished by modeling the medium by
two independent reservoirs composed of a continuum of three dimensional
harmonic oscillators. These two independent sets
of harmonic oscillators describe the polarizability and the
magnetizability characters of the MD medium and interact with the
electric and the magnetic fields through a dipole interaction
term. More details concerning the total Lagrangian density of the system is provided in Appendix~\ref{App:Details of canonical quantization}.
By defining the canonical conjugate momentums of system and making use of the Lagrangian~(\ref{total Lagrangian}), the Hamiltonian of the whole system emerging from the canonical procedure. The Hamiltonian, whose derivation can be found in Refs.~\cite{Morshed 2016,Kheirandish 2011,Amooghorban 2011}, reads in the electric dipole and the rotating-wave approximation as follows:
\begin{eqnarray}\label{Hamiltonian}
&& \hat{H}=\sum\limits_{\lambda =e,m}{\int{{{d}^{3}}r\int_{0}^{\infty }{d\omega \,\hbar \omega \text{ }\hat{\mathbf{f}}_{^{\lambda }}^{\dagger }\left( \mathbf{r},\omega  \right)}}{{{\hat{\mathbf{f}}}}_{\lambda }}\left(\mathbf{r},\omega  \right)}+\\
&& \sum\limits_{j=A,B}{\hbar {{\omega }_{j}}}{{\hat{\sigma }}_{jz}}-\sum\limits_{j=A,B}{\left[ \sigma _{j}^{\dagger }{{\mathbf{d}}_{j}}\,\cdot \int_{0}^{\infty }{d\omega \,{{{\hat{\bf{E}}}}^{(+)}}\left( {{\mathbf{r}}_{j}},\omega  \right)+H.c.} \right]}, \nonumber
\end{eqnarray}
where $\hat{{\bf f}}_e$ and $\hat{{\bf f}}_m$ denote two independent infinite sets of bosonic
operators, which are associated with the electric and magnetic
excitations of the system, ${{\sigma }_{j}}=\left| {{l}_{j}} \right\rangle \left\langle  {{u}_{j}} \right|$ and $\sigma _{j}^{\dagger }=\left| {{u}_{j}} \right\rangle \left\langle  {{l}_{j}} \right|$ are the lowering  and raising operators for
each atom, respectively, in which $\left| {{u}_{j}} \right\rangle$ and $\left| {{l}_{j}} \right\rangle $ are the upper and lower states of the atom, and ${{\bf d}_{j}}=\left\langle  {{u}_{j}} \right|{{\hat{\bf d}}_{k}}\left| {{l}_{j}} \right\rangle $ represents the dipolar transition matrix element. Furthermore, $\hat{{\bf{E}}}^+$ is the positive frequency part of the electric field operator expressed in terms of $\hat{{\bf f}}_\lambda$ and the EM Green tensor of system in Eq.~(\ref{E+}).

To investigate the entanglement dynamics of the atomic system via the spontaneous emission, we first need to obtain
the time-dependent wave function of the combined system. 
When there is only one excitation in the whole system and the EM field is initially at $t=0$ in the vacuum state, the wave function at time $t$ can be written as
\begin{eqnarray}\label{wave function}
&& \left| \psi \left( t \right) \right\rangle ={{C}_{{{u}_{A}}}}( t ) \,{{{e}^{-i({{\omega }_{A}}-\bar{\omega })t}} \left| \left\{ 0 \right\} \right  \rangle \left| {{u}_{A}} , {{l}_{B}} \right\rangle }+\nonumber\\
&&{{C}_{{{u}_{B}}}}( t )\,   {{{e}^{-i({{\omega }_{B}}-\bar{\omega })t}}  \left| \left\{ 0 \right\} \right  \rangle \left| {{l}_{A}}, {{u}_{B}} \right\rangle }+\sum\limits_{\lambda =e,m} \int d^3 r\int_0^\infty d\omega\nonumber\\
&& \times C_{\lambda ,l}( {\bf r},\omega ,t)  \,e^{-i(\omega -\bar{\omega })t}  \left| {{\bf{1}}_{\lambda}} \left( \bf{r},\omega  \right) \right\rangle \left| {{l_A}}, {{l_B}} \right\rangle ,
\end{eqnarray}
where $\bar{\omega }=\sum\limits_{j}{{{\omega }_{j}}}/2$, $\left| \left\{  0   \right\} \right\rangle$ is the vacuum state of the EM field and
$\left| {{\bf 1}_{\lambda }} \right\rangle$ is its excited state where the field is in a single quantum Fock state.
Here, the coefficients ${{C}_{{{u}_{j}}}}( t )$ are the probability amplitudes at time t for atom
$j$th being in the excited state, while the coefficient ${{C}_{\lambda ,l}} ({\bf r} ,\omega ,t  )$ gives
the probability amplitude at time to find both atoms in the ground state and a photon in the field mode $\lambda$.

Inserting the Homiltonian~(\ref{Hamiltonian}) and the wave function~(\ref{wave function}) into the Schrodinger equation and making use of the Green's tensor integral relation~(\ref{integral relation Green}), yields a set of coupled integro-differential equations for motion equations of the probability amplitudes
\begin{eqnarray}\label{motion equations of probability amplitudes}
\dot{C}_j ( t  )=\int_0^t dt' K_{jj' } ( t,t'  ){{C}_{j' }}( t' ),
\end{eqnarray}
where the memory kernel ${K_{jj'}}$ in term of the EM Green tensor of the system
are given by
\begin{eqnarray}\label{memory kernel AB}
{K_{jj'}}\left( t,{t}' \right)&=&\frac{-1}{\pi {{\varepsilon }_{0}}\hbar {{c}^{2}}}\int_{0}^{\infty }{d\omega }[{{\omega }^{2}}{{e}^{-i\left( \omega -{{\omega }_{j}} \right)t}}{{e}^{i\left( \omega -{{{\tilde{\omega }}}_{{{j}'}}} \right)\left( t-{t}' \right)}}\nonumber\\
&&\times {{\bf d}_{j}}\cdot \operatorname{Im}[\bfsfG \left( {{\mathbf{r}}_{j}},{{\mathbf{r}}_{{{j}'}}},\omega  \right)]\cdot{{\bf d}_{{{j}'}}}.
\end{eqnarray}

In order to facilitate the analysis, it is sufficient to consider the simple case that
two identical two-level atoms located at opposite positions ${\bf r}_A=-{\bf r}_B=r{\bf k}$, such that $K_{AA}(t,t')=K_{BB}(t,t')=K(t-t')$, and the atomic dipole moments are antiparallel to each other and oriented along the $z$-axis, ${\rm i.e.,}$ ${\bf d}_A=-{\bf d}_B={ d}\,\hat{\bf k}$.
Let us introduce the symmetric and antisymmetric combinations of the probability amplitudes, ${{C}_{\pm }}\left( t \right)=\left[ {{C}_{{{u}_{A}}}}\left( t \right)\pm {{C}_{{{u}_{B}}}}\left( t \right) \right]/\sqrt{2}$,
which are the collective symmetric and antisymmetric states of the two-atom system, ${\rm i.e.}$, $\left| \pm  \right\rangle =(\left| {{u}_{A}},{{l}_{B}} \right\rangle \pm \left| {l}_{A},{{u}_{B}} \right\rangle )/\sqrt{2}$. Taking into account this collective basis, we can recast Eqs.~(\ref{motion equations of probability amplitudes}) into the decoupled form
\begin{eqnarray}\label{symmetric and antisymmetric probability amplitudes}
\dot{C}_{\pm }\left( t \right)=\int_0^t dt'K_{\pm }\left( t-t' \right){{C}_{\pm }}\left( t' \right),
\end{eqnarray}
with the memory kernel
\begin{eqnarray}\label{memory kernel +-}
{K_{\pm }}(t-t')={K}(t-t')\pm K_{AB}(t-t').
\end{eqnarray}
Equations~(\ref{symmetric and antisymmetric probability amplitudes}) together with the kernel given by Eq.~(\ref{memory kernel +-}) are the basic equations for studying the time evolution of the atomic system.
We can numerically find the exact solution of the integro-differential equations~(\ref{symmetric and antisymmetric probability amplitudes}). However, there are analytical solutions in two limiting cases of the weak and strong coupling.
We concentrate our attention here on the weak coupling regime, that the Markov approximation applies. As a result of this approximation, the time integral in Eqs.~(\ref{memory kernel AB}) can replaced by a Zeta function~\cite{Morshed 2016}.
With the above approximations, we get a familiar result that the amplitudes
of the symmetric and antisymmetric states decaying exponentially with time,
\begin{eqnarray}\label{C+-}
{{C}_{\pm }}\left( t \right)=\exp\left[(-{{\Gamma }^{\pm }}/2+i{{\delta }^{\pm }})t \right]{{C}_{\pm }}\left( 0 \right),
\end{eqnarray}
where ${{\Gamma }^{\pm }}={{\Gamma }}\pm {{\Gamma }_{AB}}$ and ${{\delta }^{\pm }}={{\delta }}\pm {{\delta }_{AB}}$ are, respectively, the decay rates and level shifts of the symmetric and antisymmetric states.
Here, $\delta={{\delta }_{jj}}$ and $\Gamma= {{\Gamma }_{jj}}$ are, respectively, the Lamb shift and the spontaneous emission rate of the individual atoms, whereas ${{\delta }_{jj'}}$ and ${{\Gamma }_{jj'}}$ for $j\ne j$ represent the level shift induced by the dipole-dipole coupling and the collective damping effects, and are defined, respectively, by
\begin{eqnarray}\label{deita ij}
&& {{\delta }_{jj'}}=\frac{1}{\hbar \pi {{\varepsilon }_{0}}}P\int_{0}^{\infty }{d\omega }\frac{{{\omega }^{2}}}{{{c}^{2}}}\frac{{{\mathbf{d}}_{j}}\cdot \operatorname{Im} \left[ \bfsfG \left( {{\mathbf{r}}_{j}},{{\mathbf{r}}_{j'}},\omega  \right) \right] \cdot {{\mathbf{d}}_{j'}}}{(\omega -{{\omega }_{i}})},\,\,\,\,\,\,
\end{eqnarray}
and
\begin{eqnarray}\label{Gamma ij}
{{\Gamma }_{jj'}}=\frac{2\omega _{i}^{2}}{\hbar {{\varepsilon }_{0}}{{c}^{2}}}{{\mathbf{d}}_{j}}\cdot \operatorname{Im} \left[ \bfsfG \left( {{\mathbf{r}}_{j}},{{\mathbf{r}}_{j'}},\omega  \right) \right] \cdot {{\mathbf{d}}_{j'}}.
\end{eqnarray}
with $P$ denoting the principal value. As mentioned above, both ${\Gamma }_{jj'}$ and ${\delta }_{jj'}$ can be
evaluated from the knowledge of the EM Green's tensor of the system.
The EM Green's tensor corresponding to an ideal spherical cloak is extracted in Ref.~\cite{Morshed 2016}. A summary of some final results are given in Appendix~\ref{App:GREEN TENSOR OF THE SYSTEM}. By making use of Eq.~(\ref{scattering Green tensor}), the explicit expression for the diagonal $z$ component of the scattering part of the EM Green tensor is written as
\begin{eqnarray}\label{Green tensor zz}
\bfsfG^{\rm S}_{zz}\left( \mathbf{r},\mathbf{{r}'},\omega  \right)&=&\frac{i{{k}_{1}}{{\mu }_{0}}}{4\pi }\sum\limits_{n=0}^{\infty }{n\left( n+1 \right)}(2n+1)\nonumber\\
&&\times B_{N}^{11}{{({{z}_{n}}\left( kr \right)/kr)}^{2}}.
\end{eqnarray}
Substituting Eqs.~(\ref{Green tensor zz}) into Eqs.~(\ref{deita ij}) and~(\ref{Gamma ij}), we obtain the following analytical expressions for
both ${\Gamma }^{\pm }$  and ${\delta }^{\pm }$ which are normalized to the free space decay rate [$\Gamma_0=\frac{\omega^3 d^2}{3\hbar \pi \varepsilon_0 c^3}$]
\begin{eqnarray}
\frac{\Gamma^\pm}{\Gamma_0} &=& \frac{6\pi }{\omega } {\rm Im}\left[ \frac{i{{k}_{0}}{{\mu }_{0}}}{4\pi }\sum\limits_{n=0}^{\infty }{n\left(n+1\right)}(2n+1)\right.\\
&& \left.\times B_{N}^{11}\left(\frac{z_n(kr)}{kr}\right)^2\left(1\mp (-1)^n\right)\right],\nonumber
\end{eqnarray}
and
\begin{eqnarray}
\frac{\delta^\pm}{\Gamma_0} &=& \frac{6\pi }{\omega } {\rm Re}\left[ \frac{i{{k}_{0}}{{\mu }_{0}}}{2\pi }\sum\limits_{n=0}^{\infty }{n\left(n+1\right)}(2n+1)\right.\\
&& \left.\times B_{N}^{11}\left(\frac{z_n(kr)}{kr}\right)^2\left(1\mp (-1)^n\right)\right].\nonumber
\end{eqnarray}
%
%
\section{Entanglement dynamic of atomic subsystem}\label{Sec:Entanglement dynamic of atomic subsystem}
In this section, we study the entanglement dynamics of the two-atom system induced by the EM field interaction and mediated by the cloak. Owing to interaction with the EM field, we here encounter a mixed state rather than pure one to describe the atomic system. The entanglement in a pure state can be simply calculated using the von Neumann entropy as well as other measures, whereas for the mixed states, the measure of entanglement is more complicated. Despite a great deal of effort in past years, a global criterion for multipartite entanglement does not exist. For the simplest case of a pair
of particles, there is an elegant formula in terms of concurrence, which has been introduced by the seminal work of Hill and Wootters~\cite{Hill 1997,Wootters 1998}.

In this paper we use the concurrence, that is the widely accepted measure of entanglement, to characterize the amount of entanglement between the atoms and the entanglement dynamics. The concurrence is defined as follows:
\begin{eqnarray}\label{concurrence}
{\cal C} (t)=\max \left\{ 0,{{\lambda }_{1}}-{{\lambda }_{2}}-{{\lambda }_{3}}-{{\lambda }_{4}} \right\},
\end{eqnarray}
where the quantities ${{\lambda }_{i}}$ are the square roots of the eigenvalues
in the matrix product $\varrho=\rho_{A} \left( {{\sigma }^A_{y}}\otimes {{\sigma}^B_{y}}\right) \rho^{*}_{A} \left( {{\sigma }^A_{y}}\otimes {{\sigma }^B_{y}}\right)$ in descending order.
Here, ${\sigma }^{A/B}_{y}$ is the y component of Pauli matrix for atoms $A$ and $B$, $\rho_{A}$ is the reduced density matrix of the atomic system and $ \rho^{*}_{A}$ denoting the complex conjugate of $\rho_{A}$.
The value of the concurrence varies between 0 and 1, indicating the
separable and maximally entangled atoms, respectively.
%
\begin{figure*}[t]
\begin{minipage}[b]{0.33\linewidth}
\centering
\includegraphics[width=\textwidth]{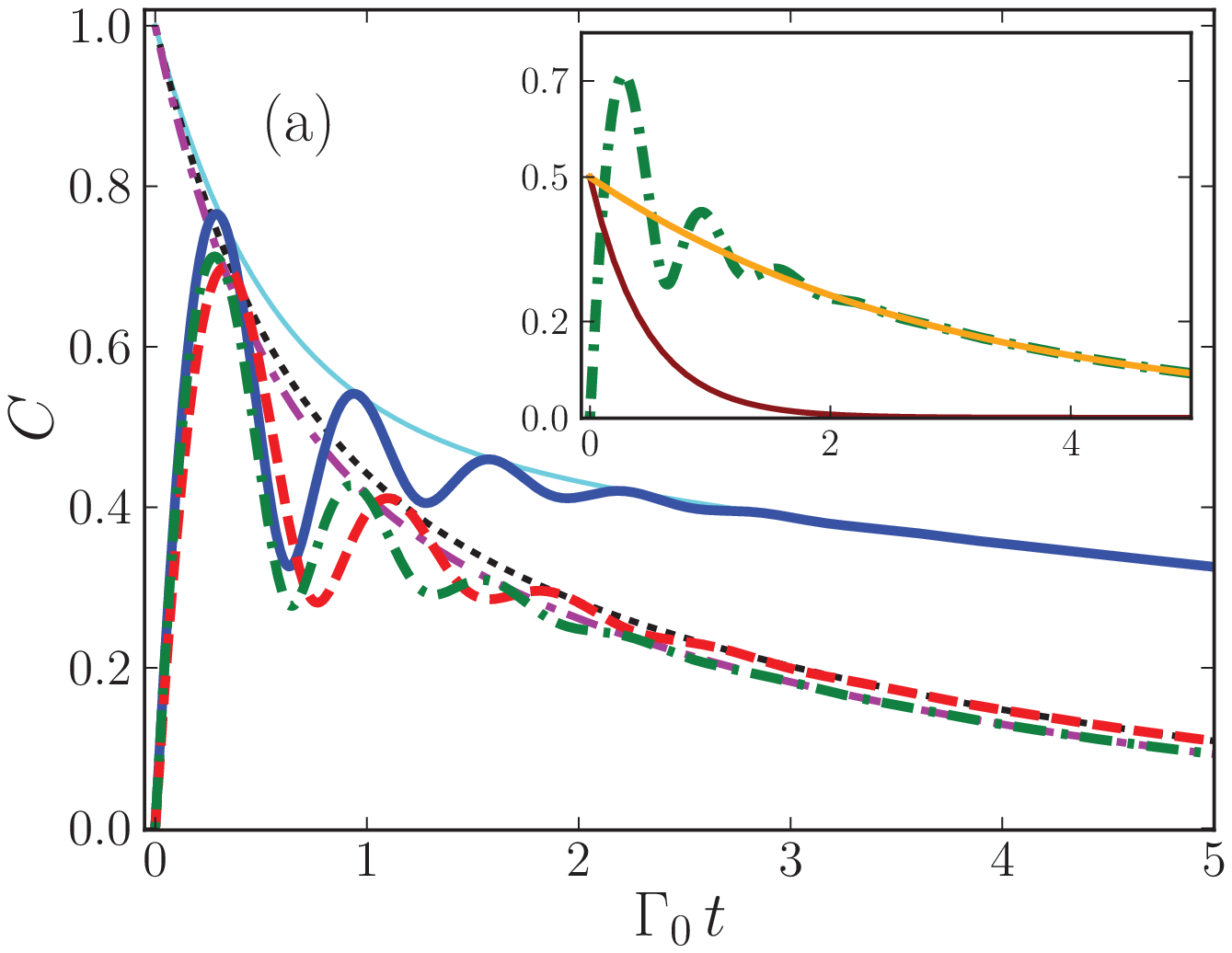}
\end{minipage}
\hspace{0cm}
\begin{minipage}[b]{0.32\linewidth}
\centering
\includegraphics[width=\textwidth]{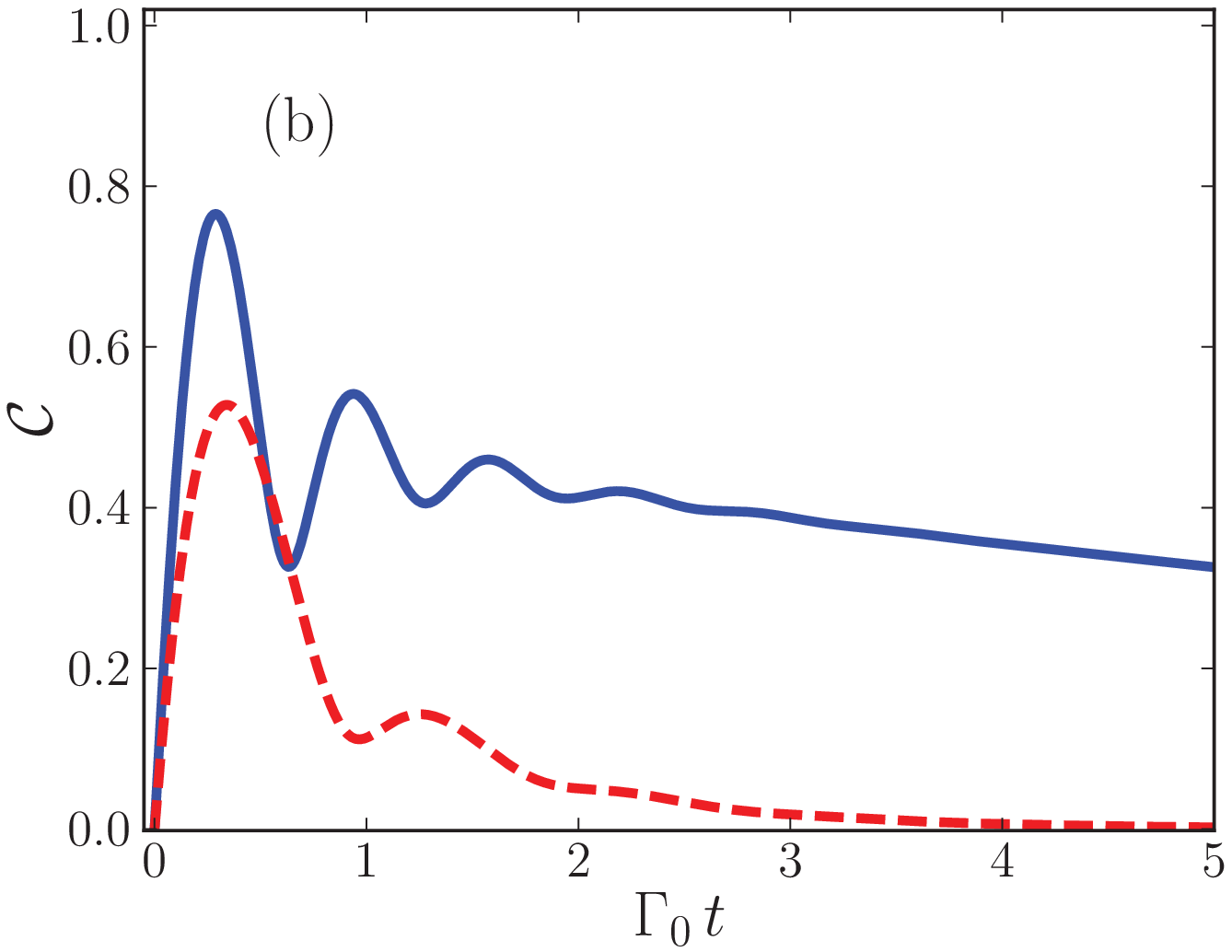}
\end{minipage}
\hspace{0cm}
\begin{minipage}[b]{0.32\linewidth}
\centering
\includegraphics[width=\textwidth]{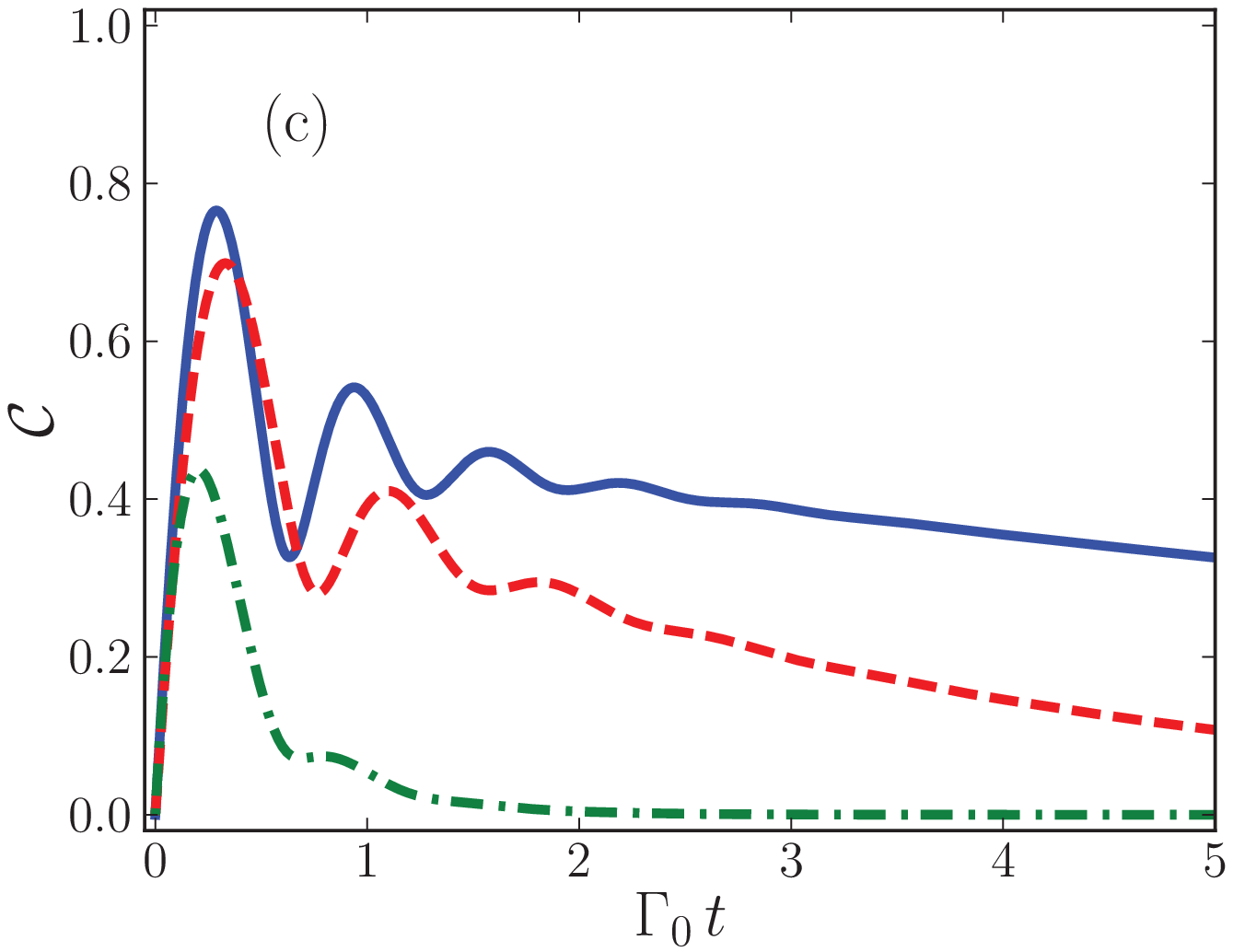}
\end{minipage}
\caption{The time evolution of the concurrence ${\cal C}$ as a function of a dimensionless parameter ${{\Gamma }_{0}}\,t$ for the case that the two-atom system is placed in free space (solid blue line), in vicinity of the clacked object (dash-dotted green line) and near the object alone (dashed red line). Time evolution of the population $|c_+|^2+|c_-|^2$ is represented by the solid cyan curve (free space), the dashed black curve (with the cloaking shell) and the dot-dashed purple line (without cloaking shell). The two identical atoms located at diametrically opposite positions $r{{\omega }_{0}}/c=4.7$. The material absorbtion of the central hidden object and the cloak are described by the Lorentz model with parameters: $\alpha =1.3$, ${{\omega }_{p}}/{{\omega }_{0}}=0.1$. In panels (a) and (b) $\gamma /\omega_0=0.01$, and in panel (c) $\gamma /\omega_0=0.1$. The inner and outer radius of the cloaking shell [panels (a) and (c)] are $b{{\omega }_{0}}/c=3$ and $ a{{\omega }_{0}}/c=4.5$, respectively, and the radius of the object without cloak in panel (b) is $b{{\omega }_{0}}/c=4.5$. The atoms are nonresonant with the cloak $\omega/\omega_0=0.1$, and initially prepared in the state $| u_A , l_B \rangle $. The inset shows the time evolution of ${\cal C}$ and the populations of states $|+ \rangle$ (solid yellow line) and $|- \rangle$ (solid brown line) in the presence of the cloaked object.}
\label{Fig:2}
\end{figure*}

%
To calculate the concurrence~(\ref{concurrence}), we first need to determine the reduced density matrix $\rho_{A}$.
For a given wave function $|\psi \rangle$, the reduced density operator of the atomic system is obtained from the total density operator $\rho =\left| \psi  \right\rangle \left\langle  \psi  \right|$, by tracing out the EM degrees of freedom. To simplify the calculations, one can work in the basis of collective states in the Hilbert space ${{\mathbb{C}}^{2}}\otimes {{\mathbb{C}}^{2}}$,
which contains the symmetric and antisymmetric states. In this basis, the two-atom system
can be treated as a single four-level system, with the ground state $| l \rangle  =| l_A ,l_B \rangle $, the upper state $| u \rangle =| u_A , u_B \rangle$, and two intermediate states $| + \rangle$ and $| -  \rangle$~\cite{Tanas 2004,Dung 2002}. The states $| + \rangle$ and $| -  \rangle$ decay to the two atom ground state $| l \rangle$	with the rates $\Gamma_+$ and $\Gamma_-$, respectively. The eigenenergies of $|\pm \rangle$ states are split by the dipole-dipole coupling.
By using the wave function given in~(\ref{wave function}) and considering the collective states, after some algebra, the reduced density operator for the two-atoms system is given by
\begin{eqnarray}
\rho_A &=& | C_+|^2 | + \rangle \langle + |   + | C_-|^2 | - \rangle \langle - |\nonumber\\
&&     +C_+ C^*_-  | + \rangle \langle - |  +C_- C^*_+  | - \rangle \langle + |\nonumber\\
&&     +(1-{{\left| {{C}_{+}} \right|}^{2}}-{{\left| {{C}_{-}} \right|}^{2}}) | l \rangle \langle l| .
\end{eqnarray}
Following the definition of the concurrence~(\ref{concurrence}) and using above equation, we find that the concurrence in terms of the populations of states $|+\rangle$ and $|-\rangle$ is written as
\begin{eqnarray}
{\cal C}( t )=\frac{1}{2} \sqrt{\left(  e^{-\Gamma_+ t} +e^{-\Gamma_- t}\right)^2 -4 e^{-2\Gamma t} \cos^2(2 \,\delta_{AB}t)}.\,\,\,\,\,\,\,\,\,\,
\end{eqnarray}
Here we assumed that the atomic system is initially prepared in the unentangled state $| u_A , l_B \rangle $, {\rm i.e.} two symmetric and antisymmetric states are initially equally excited. The first term on rhs of Eq.~(\ref{concurrence}) associated with the population sum of states $|\pm \rangle$ and exhibits an exponential relaxation of ${\cal C}$, whereas the second term represents an oscillatory behavior when the dipole-dipole coupling ${{\delta }_{jj'}}$ is sufficiently stronger than the damping rate $\Gamma$.

In what follows, we present the results of our numerical
calculations pertaining to the concurrence~(\ref{concurrence}) to study the entanglement dynamic of the two-atom system near the cloak.
For all of our calculations we take the material and geometrical parameters of the constructed cloak in Ref.~\cite{Schurig 2006} with the inner radius $b = 3c/\omega_0$ and outer radius $a = 4.5c/\omega_0$.

In Fig.~\ref{Fig:2}, the time evolution of the concurrence is plotted as a function of the dimensionless parameter ${{\Gamma }_{0}}\,t$ for two different
configurations when the atomic transition frequencies are far from the cloak resonance  $\omega =0.1{{\omega }_{0}}$ and the absorbtion is weak.
In case 1, the influence of the central hidden object alone on the entanglement of the two-atom system is represented by the dashed red line, while in case 2, the results for the combination of the cloak and object is depicted by the dot-dashed green line. For more clarification, our results are also compared to the solid blue line when the two atoms are in free space.
%
%
\
\begin{figure*}[t]
\begin{minipage}[b]{0.42\linewidth}
\centering
\includegraphics[width=\textwidth]{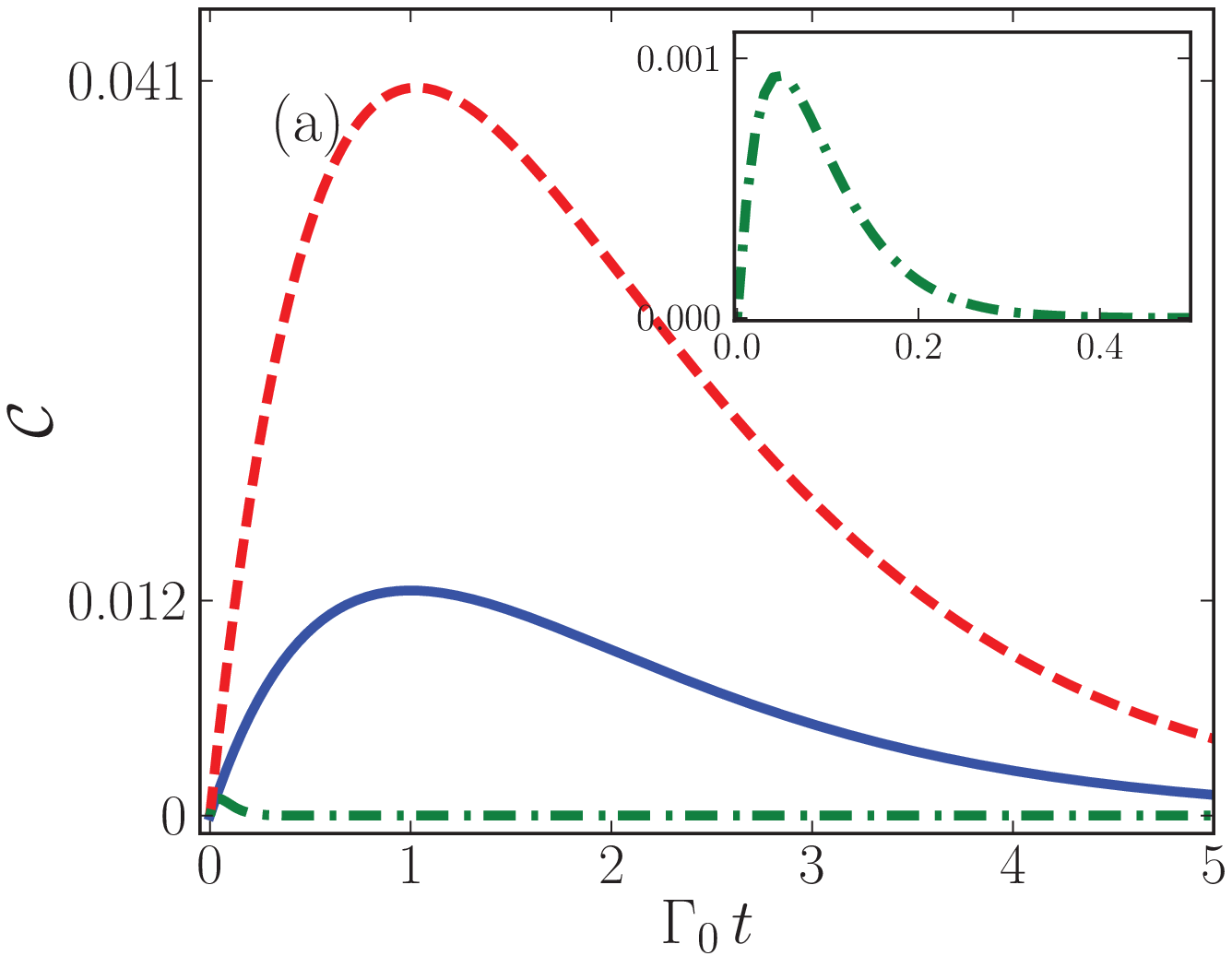}
\end{minipage}
\hspace{1cm}
\begin{minipage}[b]{0.42\linewidth}
\centering
\includegraphics[width=\textwidth]{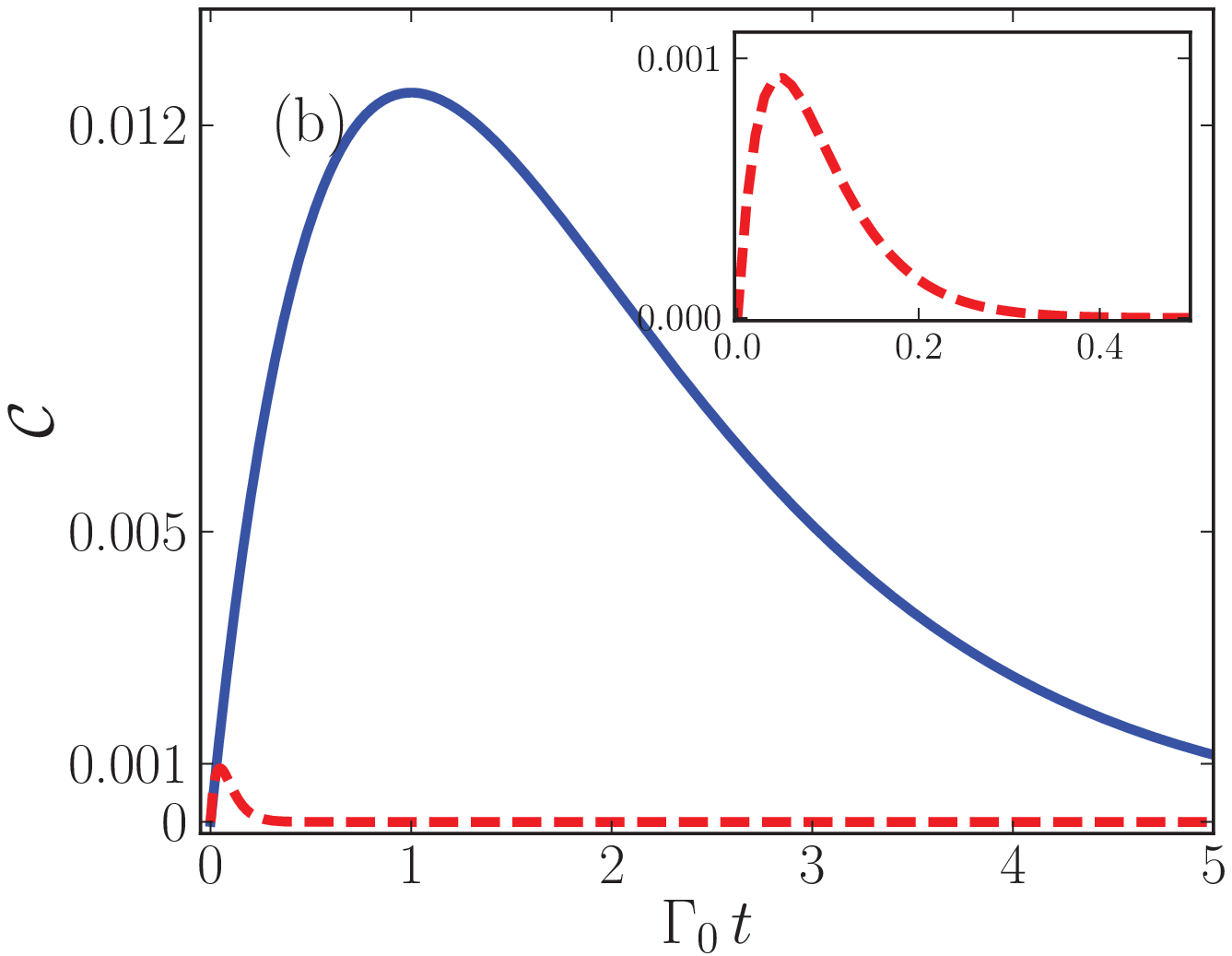}
\end{minipage}
\caption{The time evolution of the concurrence ${\cal C}$ as a function of a dimensionless parameter ${{\Gamma }_{0}}\,t$. In panel (a) the inner and outer radius of the cloaking shell is $b{{\omega }_{0}}/c=3$ and $ a{{\omega }_{0}}/c=4.5$, respectively, and in panel (b) the radius of the object without cloak is $b{{\omega }_{0}}/c=4.5$. The material parameters of the cloaking shell and the hidden object are identical to those used in Fig.~\ref{Fig:2} (a).}
\label{Fig:3}
\end{figure*}
%

In panel (a), a general behavior is observed for all curves:
we see that ${\cal C}$ is zero at the initial time $t = 0$. This is as it should be, since the initial state of the atomic system was prepared in the unentangled state $| u_A , l_B \rangle $. As time progresses, the concurrence shows an oscillatory behavior followed by a very slow decay. This oscillatory behavior, which is observed at times shorter than $2\Gamma_0^{-1}$, can be easily understood from the dominant contribution of the oscillatory term in Eq.~(\ref{concurrence}). The oscillations are with the frequency $2 \delta_{AB}$, which is equal to the separation of the symmetric and antisymmetric states.
The oscillations are also bounded from above by upper envelopes, which are given by the time evolution of the population sum $|c_+|^2+|c_-|^2$.
The time evolution of the population sum in the presence and the absence of the cloak, and in free space, respectively, are presented by the dot-dashed purple, the dotted black and the solid cyan line.
We do not depict here the time variation of the population difference $|c_+|^2-|c_-|^2$, which appear as lower envelopes for oscillations, to avoid the plots overcrowding.

After the time  $ 2\Gamma _{0}^{-1}$, the time evolution of the populations of two states $|\pm \rangle$ play a pivotal role in the evaluation of ${\cal C}$, due to the significant contribution of the first term in Eq.~(\ref{concurrence}). These states are equally populated initially, but for small atoms-cloak distance and the material parameters are used here (see caption of Fig.~\ref{Fig:2}), the values of $\Gamma_+$ and $\Gamma_-$ are different from
each other, such that the antisymmetric state (see the solid brown line in inset) decay much faster than the symmetric state (see the solid yellow line in inset). Consequently, for times longer than $2\Gamma _{0}^{-1}$,
the concurrence is only determined by the time evolution of the symmetric state, and hereby some entanglement observes for longer times. Finally, the concurrence tends to zero when the symmetric state is depopulated in the limit of very long time.
It is worth noting that the values of $\Gamma_+$ and $\Gamma_-$ at large atoms-cloak distance, not shown here, tend to the free space rate and all curves coincide on the free space case.

It can be easily discerned from Fig.~\ref{Fig:2} (a) that the effect of the cloaked object on the concurrence appears as a slight reduction in amplitude of the oscillation, so that the value of ${\cal C}$ at the first maximum is nearly $0.7$, which is lower than the corresponding value for free space equal to $0.8$, and then followed by a very fast decay as compared to free space,
These behavior comes from the modification of $\Gamma_\pm$ results from the variation of $\Gamma_{AB}$ in presence of the cloak and leads to the symmetric state decays much faster than that in the free space. Despite of this amplitude reduction, in the absence of the cloak,
the concurrence experiences a small variation on the frequency of the oscillations.
Recalling Eq.~(\ref{deita ij}), it is easy seen that the dipole-dipole coupling ${{\delta }_{AB}}$ depends on the material environment, and the distance of the two atoms through the EM Green tensor of the system. Interestingly, the time variations of ${{\delta }_{AB}}$ show no changes in the presence of the cloak and beyond its resonance frequency (not shown here), while undergoing variations in the absence of the cloak.

As can be seen when comparing the dot-dashed green line to the dotted red line in Fig.~\ref{Fig:2}(a), the amplitude reduction in this two configurations is almost the same, while the distance of the atoms from the cloak in the case 2, and from the object alone in the case 1 are not equal.
One might reasonably ask whether this behavior is a result of unusual electromagnetic properties of the cloaking material. To answer this question we have plotted in Fig.~\ref{Fig:2}(b) the time evolution of ${\cal C}$  for the case that a larger object with radius $b{{\omega }_{0}}/c=4.5$ and the same material parameters indicated in Fig.~\ref{Fig:2} (a), is placed between the two atoms.
One can see that the amplitude of the concurrence decreases with increasing the radius of the MD object and finally disappears very fast with the time evolves. This is in big contrast with the case 2 that the cloaking shell is present.
Hereby, by monitoring the entanglement dynamic of the two-atoms system at small atoms-cloak distance we can say that the performance of the cloaking device to hide the object is rather satisfactory.

In Fig.~\ref{Fig:2}(c), the results of Fig.~\ref{Fig:2}(a) have been repeated for a MD object with the absorbtion coefficient $\gamma /\omega_0=0.1$.
%
\begin{figure*}[t]
\begin{minipage}[b]{0.42\linewidth}
\centering
\includegraphics[width=\textwidth]{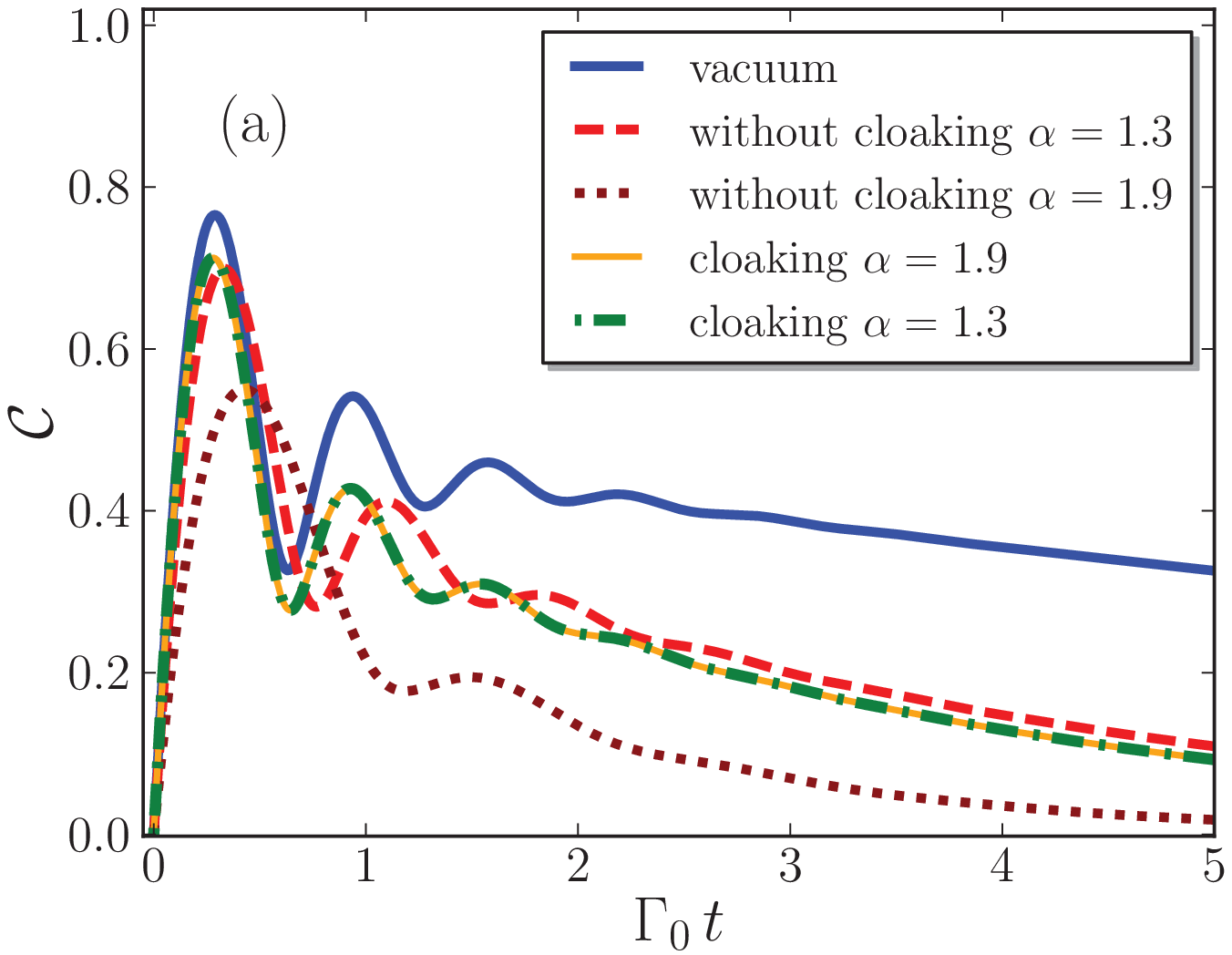}
\end{minipage}
\hspace{1cm}
\begin{minipage}[b]{0.42\linewidth}
\centering
\includegraphics[width=\textwidth]{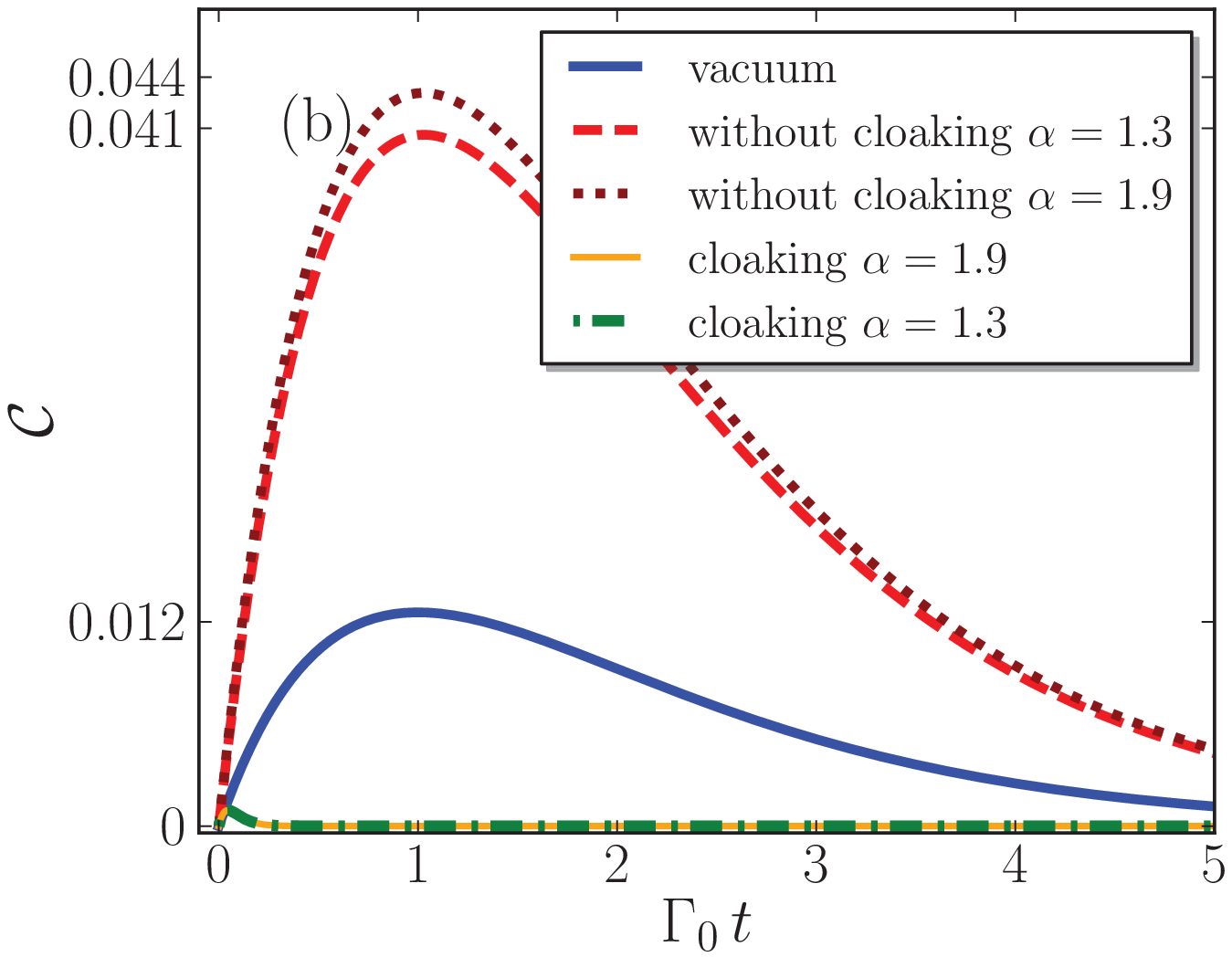}
\end{minipage}
\caption{The time evolution of the concurrence ${\cal C}$ as a function of a dimensionless parameter ${{\Gamma }_{0}}\,t$ for (a) $\omega/\omega_0=0.1$, and (b) $\omega ={{\omega }_{0}}$. The atoms are resonant with the cloak $\omega/\omega_0=1$, and initially prepared in the state $| u_A , l_B \rangle $. The geometrical and the material parameters of the cloaking shell and the hidden object are identical to those used in Fig.~\ref{Fig:2} (a).}
\label{Fig:4}
\end{figure*}
%
The time evolution of ${\cal C}$ shows a similar dynamic behavior compared to what observed in Fig.~\ref{Fig:2}(a).
With the absorbtion coefficient risen to $\gamma /\omega_0=0.1$, the effect of loss in reducing the entanglement of two atoms is perceptible, such that the value of ${\cal C}$ at the first maximum reduced to $0.4$.
No significant differences is observed when the object alone is present because the small loss plays no roll at large atoms-object distance.
Fig.~\ref{Fig:2}(c) shows that the loss degrades the cloaking performance. Although it is still effective which is unavoidable because
the transition frequency of two atoms is far from the cloak resonance and the absorption is still small.

In Fig.~\ref{Fig:3} (a), the time evolution of ${\cal C}$ is depicted as a function of a dimensionless parameter ${{\Gamma }_{0}}\,t$ for two different
configuration considered in Fig.~\ref{Fig:2}, but the two atoms are at resonance with the cloak shell $\omega =\omega_0$. In all three
cases, the concurrence reaches much lower maximum values than the nonresonance cases and then decays exponentially to zero. The
maximum values of the concurrence in these cases are $0.001$, $0.041$ and $0.012$ in the presence of the cloak shell (dash-dotted green), in the presence of the object alone (dashed red line) and in free space (solid blue line), respectively.
What is clear from these curves, there are no oscillations at short times as a consequence of the significant reduction of the dipole-dipole coupling ${{\delta }_{AB}}$ at resonance frequency.

With the high loss that occurs at resonance frequency, the cloaking shell absorbs the spontaneously emitted photon before it exits the cloak. We thus expect that
a lossy cloak would decrease the amount of entanglement created between two atoms. This is
precisely what the dash-dotted green line show. Unlike to this case, the concurrence in the absence of the cloaking shell experiences a little enhancement compared to the free space case, due to large distance of the atoms from the object.
However, it is interesting when a bigger object with a radius equal to the outer radius of the cloaking shell, $b{{\omega }_{0}}/c=4.5$, is placed between two atoms [see Fig.~\ref{Fig:3}(b)]. Similar to that case the object surrounded by the cloaking shell, some small entanglement observes on a short timescale of order $0.2\Gamma_0^{-1}$. The maximum value of the concurrence in this case is $0.001$, which is much too small due to the dissipative nature of the object.

The effect of different hidden objects inside the cloaking shell on the entanglement degree of the atomic system is illustrated in Fig.~\ref{Fig:4}.
The different objects enter to our calculations by attributing different values to the factor $\alpha$ of the material parameters.
In Fig.~\ref{Fig:4} (a), in the absence of the cloak and nonresonance case, the oscillation amplitude of ${\cal C}$ decreases and then decays very fast to zero by changing the constant factor $\alpha$ from $1.3$ to $1.9$.
For $\alpha=1.3$, the value of the concurrence at the maximum is
$0.7$, which is higher than the corresponding value for $\alpha=1.9$ equal to $0.57$.
At resonance case and without the cloak shell, the amplitude of ${\cal C}$ increases slightly and then decays exponentially to zero (see Fig.~\ref{Fig:4} (b)). In this case, the concurrence reaches maximum values of $0.041$ and $0.044$ for $\alpha=1.3$ and $\alpha=1.9$, respectively.
When the cloak shell is present, the dot-dashed green and yellow lines in both resonance and nonresonance cases show no changes with increasing the constant factor $\alpha$ from $1.3$ to $1.9$. In agreement with the classical results, this shows that the performance of the cloak shell is independent of the object to be cloaked.

\section{conclusion}\label{Sec:conclusion}
In this paper, we have examined the entanglement between two mutually independent
two-level atoms at a fixed radial distance outside an invisibility cloaking shell which surrounded a hidden object. We treat the two-atom system as a quantum
system in a bath of fluctuating quantized EM field in vacuum. We have derived an analytical expression for the probability amplitudes of the atomic states in weak coupling regime when initially only one atom is excited. On this base, the reduced density operator of the atomic system is obtained. To quantify the amount of entanglement induced by the EM field interaction and mediated by the cloaking shell, we adopted the concurrence as a good entanglement measure for $2 \times 2$ quantum states. In agreement with the classical results, we found that the effect of the cloaking on the entanglement is independent of the hidden object inside the cloak.
Sufficiently far from the cloak when the loss is small, the entanglement appears for long times. This is because of the slow decaying of the symmetric state. It is seen that the amount of entanglement in the presence of the cloak exhibits a small change in amplitude compared to the case that the atoms were in free space.
In contrast to the nonresonance case, the amount of entanglement decreases strongly due to the presence of the lossy cloaking shell. Therefore, the combination of the cloak and the object can be easily detected by monitoring the entanglement between the two atoms at resonance case.

\appendix
\section{Canonical quantization of EM field}\label{App:Details of canonical quantization}
The canonical formalism for quantization of the EM field interacting with charged particles in a lossy and dispersive MD medium has been carried out previously. We give here only the essentials needed to understand the present paper.
We start from the total Lagrangian density of the system composed of the EM field, the external charged particles, the MD medium and their interactions with each other~\cite{Morshed 2016}
\begin{eqnarray}\label{total Lagrangian}
{\cal L}&=& \frac{1}{2}\sum_{\alpha} m_{\alpha}\dot{{\bf r}}_{\alpha}^{2} + \frac{1}{2}\,\varepsilon_{0} {\dot{ \bf A}}^{2}({\bf{r}},t) -
\frac{1}{2\mu_{0}}\left(\nabla \times {\bf A}({\bf{r}},t)\right)^{2} \nonumber \\
&+&\frac{1}{2} \int_ {0}^{\infty} \mbox{d}\omega \left\lbrace {\dot {\bf X }}^{2}_{\omega}({\bf{r}},t) - \omega ^{2} {\bf X
}^{2}_{\omega}({\bf{r}},t)\right\rbrace \nonumber \\
&+& \frac{1}{2} \int_ {0}^{\infty} \mbox{d}\omega \left\lbrace {\dot {\bf Y }}^{2}_{\omega}({\bf{r}},t) - \omega ^{2} {\bf Y
}^{2}_{\omega}({\bf{r}},t)\right\rbrace  \nonumber \\
&+& \sum_{\alpha} e_{\alpha} \dot {\bf r}_{\alpha} \cdot {\bf A}
({\bf r}_{\alpha},t) + {\bf A} \cdot \dot {\bf P}({\bf{r}},t)+{\bf
M}\cdot \nabla \times  {\bf
A}({\bf{r}},t)\nonumber \\
&-& W_{coul},
\end{eqnarray}
where first term is the Lagrangian of free charged particles with particles mass $ m_{\alpha} $
and position $ {\bf r} _{\alpha} $, the second and third terms is the electromagnetic part which is expressed in term of the vector potential ${\bf A}$, the second and third lines are the electric and the magnetic parts of the MD medium which are modeled by a
continuum of harmonic oscillators with two fields $\bf{X}_{\omega} $ and $ \bf{Y}_{\omega}$, the fourth line is the interaction part
which included the linear interaction between the medium and the
charged particles with the EM field. Here, ${\bf P}$ and ${\bf M}$ are the polarization and magnetization fields of the medium, and $ W_{coul} $ is the Coulomb energy of the charged particles, the polarization-charge and their interactions.
The Lagrangian~(\ref{total Lagrangian}) can now be used to obtain the corresponding canonical conjugate variables for the dynamic variables 
\begin{subequations}\label{canonicall conjugate variables}
\begin{eqnarray}
&&\hspace{-0.5cm}- {\varepsilon _0}{\bf{E}}^{\bot}({\bf{r}},t) = \frac{{\delta L}}{{\delta \dot{\bf{A}}({\bf{r}},t)}} = {\varepsilon _0} \dot{{\bf A}}({\bf{r}},t),
\label{canonicall conjugate variables E} \\
&&\hspace{-0.5cm}{\bf Q }_\omega ({\bf{r}},t) = \frac {{\delta L}} { \delta {\dot{{\bf X}}_{\omega }({\bf r},t)}}= {\bar{\bar{g}}_{e}}\cdot{\bf{A}}({\bf{r}},t) +\dot{{\bf X }}_{\omega } ({\bf{r}},t),\label{canonicall conjugate variables Q} \\
&&\hspace{-0.5cm}{{\bf \Pi} _\omega }({\bf{r}},t) = \frac{{\delta L}} { \delta {\dot{{\bf Y }}_{\omega }({\bf r},t)}} =  {\dot{\bf{Y}}}_\omega
({\bf{r}},t),\label{canonicall conjugate variables Pi} \\
&&\hspace{-0.5cm}{\bf p}_{\alpha} = \frac{\partial
L}{\partial {\dot{r_{\alpha}}}} = m_{\alpha}{\bf r}_{\alpha} +
e_{\alpha} {\bf A}({\bf r}_{\alpha} , t )\label{canonicall conjugate
variables p}.
\end{eqnarray}
\end{subequations}
The transition from the classic to the quantum domain can be done in a standard fashion by applying commutation relation
on the variables and their corresponding conjugates. For the EM and material fields, we have
\begin{subequations}\label{commutation relation for XYp}
\begin{eqnarray}
\left[ {A_j} ({\bf r} , t ), - \varepsilon_{0}E_{j'}^{\perp} ({
{\bf r}^{\prime}} , t )\right] &=& i \hbar \delta_{jj'} \delta^{\perp}({\bf  r} - {\bf
r^{\prime}}),\\
\left[ { {X}}_{\omega j} ({\bf r} , t ), { {Q}}_{\omega^{\prime} j'} ({\bf  r^{\prime}} , t )\right] &=&  i \hbar \delta_{jj'} \delta ({\bf r} - {{\bf  r}^{\prime}}) \delta
({\omega
} - \omega^{\prime}),\hspace{0.7cm}\label{commutation relation for X} \\
\left[ { {Y}}_{\omega j} ({\bf  r} , t ), \Pi_{\omega^{\prime} j} ({\bf  r^{\prime}} , t )\right] &=&  i \hbar \delta_{jj'} \delta ({\bf r} - {{\bf r}^{\prime}}) \delta ({\omega } -
\omega^{\prime}),\label{commutation relation for Y} \\
\left[ q_{\alpha},{{ p}}_{\beta}\right] &=& i \hbar
\delta_{\alpha\beta} \label{commutation relation for p}.
\end{eqnarray}
\end{subequations}
To facilitate the calculations, we introduce the following
annihilation operators:
\begin{subequations}\label{annihilation operators}
\begin{eqnarray}
{\bf f}_{e}({\bf r},\omega , t) = \frac{1}{\sqrt{2 \hbar \omega}} \left[-i \omega {\bf X}_{\omega } ({\bf r} , t ) + {\bf Q}_{\omega } ({\bf r} , t
)\right] , \label{annihilation operatorsfe} \\
{\bf f}_{m}({\bf r},\omega , t) = \frac{1}{\sqrt{2 \hbar \omega}}
\left[ \omega {\bf Y}_{\omega } ({\bf r} , t ) + i {\bf \Pi}_{\omega
} ({\bf r} , t )\right] ,\label{annihilation operators fm}
\end{eqnarray}
\end{subequations}
where ${\bf f}_{e}$ and ${\bf f}_{m}$ denote two independent
infinite sets of bosonic operators, which associated with the electric and
magnetic excitations. By making use of Eqs.~(\ref{commutation relation for X})
and~(\ref{commutation relation for Y}), these operators satisfy the simple boson commutation relation
\begin{eqnarray}
&& \left[ {{{ \mathbf{f}}}_{\lambda j}}\left( \mathbf{r},\omega  \right), \mathbf{f}_{\lambda {j}'}^{\dagger }\left( \mathbf{{r}'},{\omega }' \right) \right]={{\delta }_{j{j}'}}\delta \left( \mathbf{r}-\mathbf{{r}'} \right)\delta \left( \omega -{\omega }' \right),\nonumber \\
&& \left[ {{{ \mathbf{f}}}_{\lambda j}}\left( \mathbf{r},\omega  \right),{{{ \mathbf{f}}}_{\lambda {j}'}}\left( \mathbf{{r}'},{\omega }' \right) \right]=0.
\end{eqnarray}

By applying the Lagrangian~(\ref{total Lagrangian}) and the canonical conjugate variables~(\ref{canonicall conjugate variables}), we can form the Hamiltonian of the system. In the electric-dipole and rotating wave approximation, the Hamiltonian of the system is reduced to Eq.~(\ref{Hamiltonian}).
The Heisenberg equations of motion for dynamical variables of the system are obtained  by making use of the Hamiltonian~(\ref{Hamiltonian}) and the commutation relations~(\ref{commutation relation for XYp}). These equations for the vector potential and the transverse electric
field lead to a quantum version of the Maxwell equations and Lorentz-force law~\cite{Kheirandish 2010,Amooghorban 2011,Morshed 2016}. With the usual decompositions of the fields into positive and negative frequency components and the frequency-space Fourier transform of the fields, the Maxwell's equations in combination with the constitutive equations reduce to the wave equation
\begin{eqnarray}\label{wave equations}
&& \nabla \times  \bar{\bar{\boldsymbol \mu}}^{-1}\left( \mathbf{r},\omega  \right)\nabla \times {{{\mathbf{E}}}^{(+)}}\left(\mathbf{r},\omega  \right)\nonumber\\
&& -\frac{{{\omega }^{2}}}{{{c}^{2}}}\bar{\bar{\boldsymbol \varepsilon }}\left( \mathbf{r},\omega  \right){{{\mathbf{E}}}^{(+)}}\left( \mathbf{r},\omega  \right)=i\omega {{\mu }_{0}}{{{\mathbf{j}}}^{N(+)}}\left(\mathbf{r},\omega  \right),
\end{eqnarray}
where $\bar{\bar{\boldsymbol \varepsilon }}$ and $ \bar{\bar{\boldsymbol \mu}}$ are the electric permittivity and the magnetic permeability tensors of the medium, respectively, and ${{{\mathbf{j}}}^{N(+)}}$ is the positive frequency part of the noise current density associated with the noise sources in the MD medium. The noise current density operator  ${{{\mathbf{j}}}^{N(+)}}$ is related to the noise polarization and the noise magnetization operators as
\begin{eqnarray}
{{{\mathbf{j}}}^{N(+)}}\left({\bf r},\omega  \right)=-i\omega {{{\mathbf{P}}}^{N(+)}}\left(\mathbf{r},\omega  \right)+\nabla \times {{{\mathbf{M}}}^{N(+)}}\left({\bf r},\omega  \right),\nonumber\\
\end{eqnarray}
where 
\begin{eqnarray}
{{{\mathbf{P}}}^{N(+)}}(\mathbf{r},t)=i\int_{0}^{\infty }{d\omega \sqrt{\frac{\hbar }{2\omega }}\, \bar{\bar{g}}_e (\mathbf{r},\omega )\cdot {{{ \mathbf{f}}}_{e}}\left( \mathbf{r},\omega ,0 \right){{e}^{-i\omega t}}},\nonumber\\
{{{\mathbf{M}}}^{N(+)}}(\mathbf{r},t)=\int_{0}^{\infty }{d\omega \sqrt{\frac{h}{2\omega }} \, \bar{\bar{g}}_m (\mathbf{r},\omega )\cdot{{{ \mathbf{f}}}_{m}}\left( \mathbf{r},\omega ,0 \right){{e}^{-i\omega t}}}.\nonumber\\
\end{eqnarray}

The solution of Eq.~(\ref{wave equations}) for the electric field operator
is obtained by standard Green tensor methods in the form
\begin{eqnarray}\label{E+}
{\bf E}^{(+)}\left(\mathbf{r},\omega  \right)=i\omega {{\mu }_{0}}\int{{{\mbox d}^{3}}}r\, \bfsfG \left( \mathbf{r},\mathbf{r'},\omega\right)\cdot {{{\mathbf{j}}}^{N(+)}}\left(\mathbf{r'},\omega  \right),\,\,\,\,\,\,\,\,
\end{eqnarray}
where $\bfsfG ( \mathbf{r},\mathbf{r'},\omega )$ is the
EM Green tensor satisfying the Helmholtz equation
\begin{eqnarray}
&& \nabla \times  \bar{\bar{\boldsymbol \mu}}^{-1}\left( \mathbf{r},\omega  \right)\nabla \times \,\bfsfG\left( \mathbf{r},\mathbf{r'},\omega  \right) \nonumber\\
&& -\frac{\omega^2}{c^2}\bar{\bar{\boldsymbol \varepsilon }}\left(\mathbf{r},\omega  \right)\bfsfG\left( \mathbf{r},\mathbf{r'},\omega  \right)=\delta \left( \mathbf{r}-\mathbf{r'} \right)\bar{\bar{\bf I}},
\end{eqnarray}
together with the boundary condition $\bfsfG ({\bf r}',{\bf r},\omega)\rightarrow 0$ for $|{\bf r}-{\bf r}'|\rightarrow \infty$. In particular, the Green tensor satisfies the integral relation~\cite{Dung 2003}
\begin{eqnarray}\label{integral relation Green}
&&\int {\mbox d}^3 s\, {\rm Im}\left[\bar{\bar{\boldsymbol \mu}}^{-1}({\bf s},\omega)\right]\left[\bfsfG ({\bf r},{\bf s},\omega)\times \overleftarrow{\nabla}_s\right]\left[\overrightarrow{\nabla}_s \times \bfsfG^* ({\bf s},{\bf r}',\omega) \right] \nonumber\\
&&+\frac{\omega^2}{c^2} {\rm Im}\left[\bar{\bar{\boldsymbol \varepsilon}}({\bf s},\omega)\right] \bfsfG ({\bf r},{\bf s},\omega) \bfsfG^* ({\bf s},{\bf r}',\omega) ={\rm Im}\left[\bfsfG ({\bf r},{\bf r}',\omega)\right].\nonumber\\
\end{eqnarray}

\section{ Green tensor of the system}\label{App:GREEN TENSOR OF THE SYSTEM}
The calculation of the EM Green's tensor of an ideal spherical cloaking shell which enclosed an arbitrary object has been carried out previously in Ref.\cite{Morshed 2016}.
With regards to the fact that the two-atom system is located in free space outside the cloaking shell, the observation
point ${\bf r}$ and the source point ${\bf r}'$ are also outside the cloaking shell. Therefore, based on the method of scattering superposition, the EM Green tensor of the system outside the cloaking shell can be decomposed into two parts
\begin{eqnarray}
\bfsfG\left( \mathbf{r},\mathbf{{r}'},\omega  \right)=\bfsfG{^{\rm V}}\left( \mathbf{r},\mathbf{{r}'},\omega  \right)+\bfsfG {^{\rm S}}\left( \mathbf{r},\mathbf{{r}'},\omega  \right),
\end{eqnarray}
where $\bfsfG{^{\rm V}}$ and $\bfsfG{^{\rm S}}$ are, respectively, the vacuum and scattering contribution of the Green tensor. These two contributions under the spherical coordinate system can be expressed as follows:
\begin{eqnarray}\label{Gv}
\bfsfG{^{\rm V}}\left( {{\bf{r}},{\bf{r'}},\omega } \right)
=\frac{- \hat{\bf r}\hat{\bf r}}{{\omega}^{2}\varepsilon_{0}}\delta(r-r^{\prime})+\frac{ik_{1}\mu_{0}}{4\pi} \sum_{n=0}^{\infty}\sum_{m=0}^{\infty}
D_{m,n}\nonumber \\
 \hspace{-0.5cm}\times  \left\lbrace  \begin{array}{rl} {M_{_o^emn}^{(1)}({k_1}){\mkern 1mu} M_{_o^emn}^{\prime (1)}({k_1})} + {N_{_o^emn}^{(1)}({k_1}){\mkern 1mu} N_{_o^emn}^{\prime (1)}({k_1})}, \\
r \geq r^{\prime},\nonumber \\
{M_{_o^emn}({k_1}){\mkern 1mu} M_{_o^emn}^{\prime (1)}({k_1})} + {N_{_o^emn}({k_1}){\mkern 1mu} N_{_o^emn}^{\prime (1)}({k_1})},\\
 r \leq r^{\prime}, \nonumber \\
\end{array} \right. ,\\
\end{eqnarray}
and
\begin{eqnarray}\label{scattering Green tensor}
&&\bfsfG{^{\rm S}} \left( \mathbf{r},\mathbf{{r}'},\omega  \right)=\frac{i{{k}_{1}}}{4\pi }\sum\limits_{n=0}^{\infty }{\sum\limits_{m=0}^{n}{\left( 2-\delta _{m}^{0} \right)}}\frac{2n+1}{n\left( n+1 \right)}\frac{\left( n-m \right)!}{\left( n+m \right)!}\nonumber\\
&&\times \left[ B_{M}^{11} \,{{\mathbf{M}}}_{_{o}^{e}mn}^{\left( 1 \right)}\left( {{k}_{1}} \right) {\mathbf{{M}'}}_{_{o}^{e}mn}^{\left( 1 \right)}\left( {{k}_{1}} \right)
+B_{N}^{11}\, {{\mathbf{N}}}_{_{o}^{e}mn}^{\left( 1 \right)}\left( {{k}_{1}} \right) {\mathbf{{N}'}}_{_{o}^{e}mn}^{\left( 1 \right)}\left( {{k}_{1}} \right)\right].\nonumber \\
\end{eqnarray}
Here ${{\mathbf{M}}}_{{}_{o}^{e}mn}^{\left( 1 \right)}\left( {{k}_{1}} \right)$ and ${{\mathbf{M}}}_{{}_{o}^{e}mn}^{\left( 1 \right)}\left( {{k}_{1}} \right)$ are the spherical vector wave functions and defined as
\begin{eqnarray}
{{\mathbf{M}}}_{{}_{o}^{e}mn}^{\left( 1 \right)}\left( {{k}_{1}} \right)&=&\mp \frac{m}{\sin \theta }{{z}_{\nu }}\left( k_1r \right)P_{n}^{m}\left( \cos \theta  \right)\left( \begin{matrix}
   \operatorname{\sin}m\phi   \\
   \operatorname{\cos}m\phi   \\
\end{matrix} \right)\hat{\boldsymbol \theta }\nonumber\\
&&-{{z}_{\nu }}\left( k_1r \right)\frac{dP_{n}^{m}\left( \cos \theta  \right)}{d\theta }\left( \begin{matrix}
   \operatorname{\cos}m\phi   \\
   \operatorname{\sin}m\phi   \\
\end{matrix} \right)\hat{\boldsymbol \phi },\nonumber
\end{eqnarray}
\begin{eqnarray}
{{\mathbf{N}}}_{{}_{o}^{e}mn}^{\left( 1 \right)}\left( {{k}_{1}} \right)&=&\frac{\nu (\nu +1)}{k_1r}{{z}_{\nu }}\left( kr \right)P_{n}^{m}\left( \cos \theta  \right)\left( \begin{matrix}
   \cos m\phi   \\
   \sin m\phi   \\
\end{matrix} \right)\text{ }\hat{ \boldsymbol r} \nonumber\\
&&\hspace{-1cm}+\frac{1}{k_1r}\frac{d\left[ r{{z}_{\nu }}\left( k_1r \right) \right]}{dr}\frac{dP_{n}^{m}\left( \cos \theta  \right)}{d\theta }\left( \begin{matrix}
   \cos m\phi   \\
   \sin m\phi   \\
\end{matrix} \right)\text{ }\hat{\boldsymbol\theta }\nonumber \\
&&\hspace{-1cm}\mp \frac{1}{k_1r}\frac{d\left[ r{{z}_{\nu }}\left( k_1r \right) \right]}{dr}\frac{m}{\sin \theta }P_{n}^{m}\left( \cos \theta  \right)\left( \begin{matrix}
   \sin m\phi   \\
   \cos m\phi   \\
\end{matrix} \right) \hat{\boldsymbol \phi }\nonumber,
\end{eqnarray}
with $\nu =\sqrt{n(n+1){{\varepsilon }_{t}}/{{\varepsilon }_{r}}+1/4}-1/2$, ${{k}_{1}}=\omega/c$ and ${{k}_{t}}=\omega \sqrt{{{\varepsilon }_{t}}{{\mu }_{t}}}/c$. By imposing the boundary conditions at the surface of the cloaking shell, the unknown coefficients in Eq.~(\ref{scattering Green tensor}) are given by

\begin{eqnarray}\label{B MN}
B_{M,N}^{11}=-\frac{T_{F1}^{H,V}R_{F1}^{H,V}+T_{F1}^{H,V}R_{F2}^{H,V}}{T_{F1}^{H,V}+T_{P1}^{H,V}},
\end{eqnarray}
where the superscript $F$ and $P$ stand for the centrifugal and
centripetal waves, and the $TE$ and $ TM$ waves are represented
by the subscripts H and V , respectively. The reflection and transition coefficients $R^{H,V}_{F,P}$ and $T^{H,V}_{F,P}$ introduced in Eq.~(\ref{B MN}) are defined as
\begin{eqnarray}\label{R&T}
T_{F1}^H &=& \frac{{{\eta _1}{\mu _1}{k_2}\left( {\partial {\Im
_2}{\hbar _2} - {\Im _2}\partial {\hbar _2}} \right)}}{{{\mu
_1}{k_2}\partial {\Im _2}{\hbar _1} - {\mu _2}{k_1}{\Im _2}\partial
{\hbar _1}}}, \nonumber \\
 T_{F1}^V &=& \frac{{{\eta _1}{\mu
_1}{k_2}\left( {{\Im _2}\partial {\hbar _2} - \partial {\Im
_2}{\hbar _2}} \right)}}{{{\mu _1}{k_2}{\Im _2}\partial {\hbar _1} -
{\mu _2}{k_1}\partial {\Im _2}{\hbar _1}}},\nonumber \\
T_{P1}^V
&=& \frac{{{\eta _1}{\mu _1}{k_2}\left( {\partial {\Im _2}{\hbar _2}
- {\Im _2}\partial {\hbar _2}} \right)}}{{{\mu _1}{k_2}\partial {\Im
_1}{\hbar _2} - {\mu _2}{k_1}{\Im _1}\partial {\hbar _2}}}, \nonumber \\
T_{P1}^H &=& \frac{{{\eta _1}{\mu _1}{k_2}\left( {\partial {\Im _2}
{\hbar _2} - \partial {\Im _2}{\hbar _2}} \right)}}{{{\mu
_1}{k_2}{\Im _1}\partial {\hbar _2} - {\mu _2}{k_1}\partial {\Im
_1}{\hbar _2}}},\\
R_{F2}^H &=& \frac{{{\mu
_2}{k_3}\partial {\Im _3}{\Im _2} - {\mu _3}{k_2}\partial {\Im
_2}{\Im _3}}}{{{\mu _2}{k_3}\partial {\Im _3}{\hbar _2} - {\mu
_3}{k_2}{\Im _3}\partial {\hbar _2}}}, \nonumber \\
 R_{F2}^V &=& \frac{{{\mu
_2}{k_3}{\Im _3}\partial {\Im _2} - {\mu _3}{k_2}{\Im _2}\partial
{\Im _3}}}{{{\mu _2}{k_3}{\Im _3}\partial {\hbar _2} - {\mu
_3}{k_2}\partial {\Im _3}{\hbar _2}}},\nonumber \\
R_{F1}^V &=&
\frac{{{\mu _1}{k_2}{\Im _2}\partial {\Im _1} - {\mu _2}{k_1}{\Im
_1}\partial {\Im _2}}}{{{\mu _1}{k_2}{\Im _2}\partial {\hbar _1} -
{\mu _2}{k_1}\partial {\Im _2}{\hbar _1}}}, \nonumber \\
R_{F1}^H &=&
\frac{{{\mu _1}{k_2}\partial {\Im _2}{\Im _1} - {\mu
_2}{k_1}\partial {\Im _2}{\Im _1}}}{{{\mu _1}{k_2}\partial {\Im
_2}{\hbar _1} - {\mu _2}{k_1}{\Im _2}\partial {\hbar _1}}}.\nonumber 
\end{eqnarray}
Here, the following parameters have been used in Eq.~(\ref{R&T}) to
simplify the symbolic calculations
\begin{eqnarray}\label{hankel cl}
{\Im _{1,3}} &=& {j_{n}}\left( {{k_{1,3}}{r}} \right),\nonumber\\
{\Im _{2}} &=& {j_{n}}( {k_t}{({r-b)}}),\nonumber \\ 
{\hbar _{1}} &=& h_n^{\left( 1 \right)}\left( {{k_t}{r}} \right),\nonumber\\
{\hbar _{2}} &=& h_n^{\left( 1 \right)}\left( {k_t}{({r-b)}} \right),\nonumber\\
\partial {\Im_{1,3}} &=& \frac{1}{\rho }\frac{{\mbox{d}\left[ {\rho {j_n}\left( \rho  \right)} \right]}}{{\mbox{d}\rho }}{|_{\rho  = {k_{1,3}}{r}}}, \nonumber\\
\partial {\Im _{2}} &=& \frac{1}{\rho }\frac{{\mbox{d}\left[ {\rho {j_n}\left( \rho  \right)} \right]}}{{\mbox{d}\rho }}{|_{\rho  = {k_t}({r-b)}}},\nonumber\\
\qquad\partial{\hbar _{1}} &=& \frac{1}{\rho }\frac{{\mbox{d}\left[ {\rho h_n^{\left( 1 \right)}\left( \rho  \right)}
\right]}}{{\mbox{d}\rho }}{|_{\rho  = {k_1}{r}}},\nonumber\\
\qquad\partial{\hbar_{2}} &=& \frac{1}{\rho }\frac{{\mbox{d}\left[ {\rho h_n^{\left( 1 \right)}\left( \rho  \right)} \right]}} {{\mbox{d}\rho }}{|_{\rho  =
{k_t}({r-b)}}},\nonumber
\end{eqnarray}
where $j_n$ and $h_{n}^{(1)}$ are, respectively, the spherical Bessel function and the spherical Hankel function of the first kind, and we have the subscript $1,3$ and $2$ for outside the cloaking shell, inside the object, and inside the cloaking shell, respectively.
%

\end{document}